\documentclass[preprint,
superscriptaddress,
frontmatterverbose,
amsmath,amssymb,
aps,
prl,
]{revtex4-1}

\usepackage{graphicx}% Include ure files
\usepackage{dcolumn}% Align table columns on decimal point
\usepackage{bm}% bold math
\usepackage{verbatim}
\usepackage{hyperref}% add hypertext capabilities
\usepackage{amsmath}
\usepackage{multirow}
\usepackage{siunitx}

\usepackage{ulem}
\usepackage{color}
\begin{document}

\preprint{}

\title{Fixing the Energy Scale in Scanning Tunneling Microscopy on Semiconductor Surfaces}
%\title{TITEL}

\author{Gerhard M\"unnich}
\affiliation{Institute of Experimental and Applied Physics, University of Regensburg, 93053 Regensburg, Germany}
%\email{gerhard.muennich@physik.uni-regensburg.de}
\author{Andrea Donarini}
\affiliation{Institute of Theoretical Physics, University of Regensburg, 93053 Regensburg, Germany}
\author{Martin Wenderoth}
\affiliation{ IV. Physikalisches Institut der Georg-August-Universit\"at G\"ottingen, Friedrich-Hund-Platz 1, 37077 G\"ottingen, Germany}
\author{Jascha Repp}
\affiliation{Institute of Experimental and Applied Physics, University of Regensburg, 93053 Regensburg, Germany}

%\affiliation{Institute of Experimental and Applied Physics, University of Regensburg, 93053 Regensburg, Germany}

\date{\today}% It is always \today, today,

\begin{abstract}
In scanning tunneling experiments on semiconductor surfaces, the energy scale within the
tunneling junction is usually unknown due to tip-induced band bending. Here, we experimentally
recover the zero point of the energy scale by combining scanning tunneling microscopy with Kelvin
probe force spectroscopy. With this technique, we revisit shallow acceptors buried in GaAs.
Enhanced acceptor-related conductance is observed in negative, zero, and positive band-bending regimes.
An Anderson-Hubbard model is used to rationalize our findings, capturing the crossover between the acceptor
state being part of an impurity band for zero band bending, and the acceptor state being split off and localized for strong negative
or positive band bending, respectively.
\end{abstract}



\maketitle

%---------------------------\section{Introduction}------------------------------
Since its invention, the scanning tunneling microscope (STM)
has been widely used to study semiconductor surfaces.
The qualitative interpretation of such
studies can be obscured by the presence of tip-induced band bending TIBB($V$),
i.e.,~by the bias-dependent shift of all electronic states beneath the
microscope's tip~\cite{Feenstra87, McEllistrem93, Garleff11}.
If shifted across the Fermi level, TIBB($V$) changes the average occupation of an electronic state, which, in turn,
determines if this state contributes to the electronic transport within the junction~\cite{Marczinowski08, Teichmann08, Lee10, Zheng11, Schofield13}.
However, since the contact potential difference (CPD) between tip and sample is,
with few exceptions~\cite{Morgenstern98, Morgenstern08}, unknown in STM,
within the relevant bias range, not even the sign of TIBB($V$) is known.
In this context, the conductance spectra of shallow acceptors buried
in III--V semiconductor hosts
remained a puzzle unsolved for almost two decades:
depending on the sign of the band bending assumed or
inferred from scanning tunneling spectroscopy (STS),
conductance is explained either due to tunneling of electrons into empty
acceptor states (positive TIBB)~\cite{Mahieu05},
or due to a modification of the tunneling barrier by the occupied acceptor (negative TIBB)~\cite{Loth06},
or by the empty acceptor state being in resonance with an impurity band (zero TIBB)~\cite{Wijnheijmer10}.
Although the need for an exact value of the CPD has clearly been recognized~\cite{Loth06, Wijnheijmer10},
bare STM-based methods used so far seem not to be sufficient to resolve this puzzle.

To this end, we combine STM on a semiconducting surface with
Kelvin probe force spectroscopy (KPFS)~\cite{Nonnenmacher91},
which allows an independent and direct measurement of the CPD,
which fixes  the polarity of TIBB($V$) for all voltages.
With this combination, we revisit the shallow acceptor Zn buried in GaAs~\cite{Zheng94, Mahieu05, Loth06, Wijnheijmer10, deKort01}.
Our method reveals that the enhanced conductance induced by shallow acceptors
is not only present in one single band-bending regime,
as argued in previous publications~\cite{Mahieu05, Loth06, Wijnheijmer10},
but {\it similarly} in the regimes of negative, zero, and positive TIBB($V$).

The spatially localized band bending in an STM setup will
split off the foremost acceptor state from an
impurity band. In the most simple picture~\cite{Mahieu05, Loth06, Wijnheijmer10},
this state was treated as being isolated.
However, for small TIBB$(V)$,  this state is still part of the delocalized impurity band,
and only with increasing TIBB$(V)$
is it gradually becoming split off and isolated.
The increasing localization of the state will affect its
charging energy, which may become relevant when considering charge transport.~%
Hence, not only TIBB$(V)$ itself but also the {\it effective} electronic coupling and the charging energy
of the acceptor state change with increasing bias voltage during spectra acquisition.
Additionally, the band bending will affect more than just a single acceptor.
Accordingly, we treat the electronic transport within the junction
using an Anderson-Hubbard model for the foremost acceptor
states which are affected by TIBB$(V)$.

%------------------------------------------\section{Experimental}---------------------------------------
\begin{figure}[]
\includegraphics[]{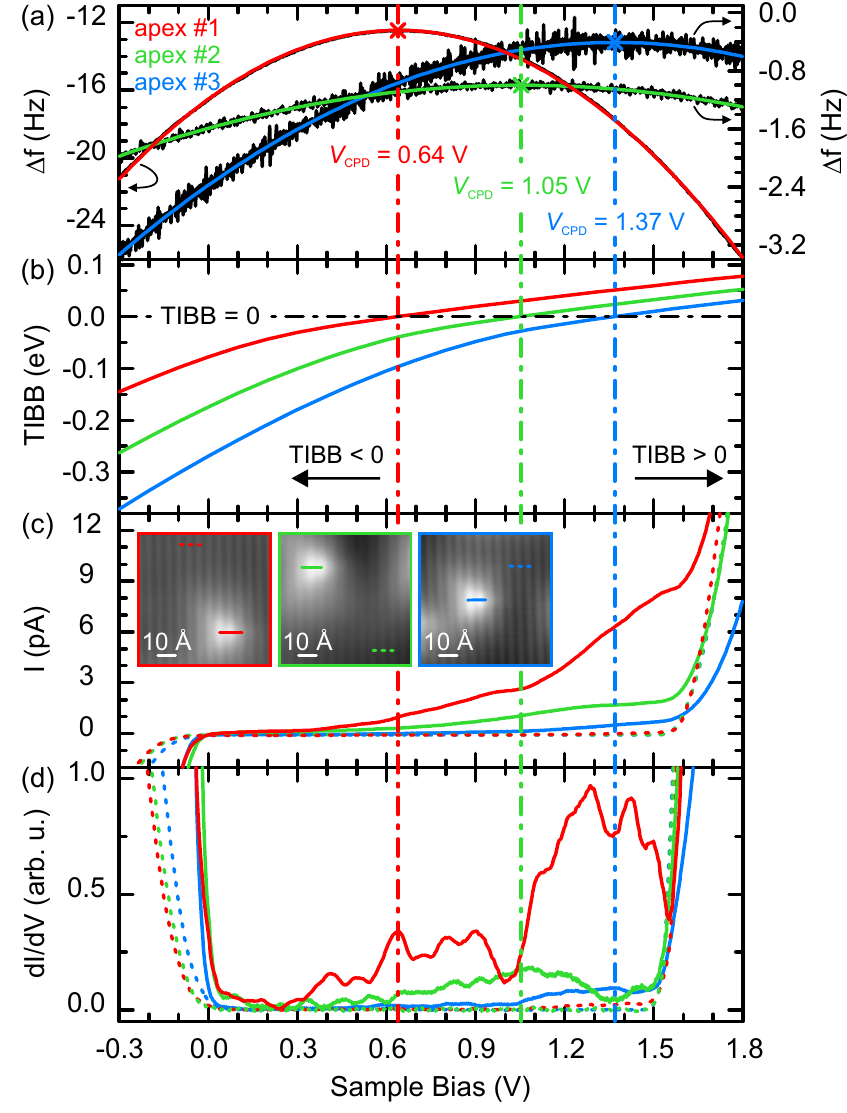}
\caption{
            Contact potential difference and
            acceptor-induced conductance for different tip apices.
            (a)~Frequency shifts measured as a function of sample bias $\Delta f(V)$
            (black lines), parabolic fits, and the corresponding flat-band voltages $V_{\textrm{CPD}}$ are indicated (colored lines).
            (b)~Calculated tip-induced band bending.
	   For any sample bias below (above) $V_{\textrm{CPD}}$,
            the band bending is negative (positive).
            (c),(d)~show $I(V)$ and $dI/dV(V)$ spectra
            away from (dashed lines) and atop (solid lines)
            sub-surface acceptors, positions are indicated in the
            constant-current STM images (inset:~$V=1.5$~V, $I=20$~pA).
	   For all tip apices, acceptor-induced enhanced current and
            conductance are observed in negative, zero, and positive
	   band-bending regimes.
}
\end{figure}

Experiments were performed by means of a
combined STM and atomic force microscope, which
was operated in ultrahigh vacuum
at a temperature of $\sim$5~K, using a qPlus force sensor
equipped with a Pt/Ir tip~\cite{Giessibl00}.
As samples, we use commercially available
GaAs wafers, which are cleaved
{\it in situ} to expose the (110) surface.
The samples are {\it p} type doped with Zn,
at an average dopant concentration of $1\times 10^{19}$~cm$^{-3}$,
which establishes an impurity band of  24~meV width,
centered around 31~meV above the valence-band edge~\cite{bookSchubert}.
This concentration corresponds to an average nearest-neighbor acceptor distance of $\approx$ 45~\AA\
and a penetration depth of the field of roughly twice this length~\cite{Feenstra03}.
Assuming a tip radius of $\gtrsim 150$~\AA, we expect about ten acceptors
to be located within the TIBB-induced space-charge region~\cite{SOM}.
%---------------------------------\section{Extracting the band bending situation from KPFS}------------------------------------

Figure~1 shows STM and KPFS measurements performed with three different tip apices.
These have been changed by controlled indentation into a clean Cu(111) surface. %~\cite{SOM}.
In the following, we will discuss data acquired with tip apex $\#1$ (red lines in Fig.~1)
while showing the results for three different tip apices to underscore the general validity of our findings.
Figure~1(a) shows KPFS data.
In KPFS, the frequency shift $\Delta f(V)$ of the force sensor is recorded  as a function of the dc sample
bias $V$ at a fixed tip position.
The electrostatic contribution to the force between tip and sample
gives rise to a parabolic dependence of $\Delta f(V)$
with $V$ as \mbox{$\Delta f(V)\propto-(V-V_\textrm{CPD})^2$}~\cite{bookSadewasser}.
For compensated CPD, that is, for $V=V_\textrm{CPD}$,
the electrostatic field in the tip-sample junction will be zero
and $\Delta f(V)$ will be maximal, respectively.
In this situation,
there is no electric field to penetrate the semiconductor and hence
$V_\textrm{CPD}$ is the flat-band voltage~\cite{Rosenwaks04}.
From the parabolic fit to $\Delta f(V)$ [cf.~Fig~1(a)],
we extract a flat-band voltage of $+0.64$~V for tip apex \#1~\cite{errorKPFM}.
The assignment of $V_\textrm{CPD}$ to the flat-band condition
relies on the GaAs(110) surface not being subject to Fermi-level pinning,
our cleaved surface being atomically flat, and our sample being
homogeneous and well conducting at 5\,K.
As this assignment is the key to our experiments,
its uncertainty was quantified as follows.
(i) Performing KPFS on GaAs at a set of different tip-sample distances
showed that local variations of the work function~\cite{Krok08} of the tip apices used
contribute to this uncertainty only by about $\pm30$~meV, and
(ii) we measured $V_\textrm{CPD}$ values on GaAs(110)
and on clean Cu(111) with the same tip apices and compared the
differences to the values expected from literature~\cite{SOM}.
This provides a generous upper bound for the uncertainty of the absolute value
of $V_\textrm{CPD}$ of $0.12$~eV~\cite{Sommerhalter99, Melitz10},
which is still small compared to the voltage scales considered here.
Finally, we note that tunneling current vs~tip-sample distance $[I(z)]$ spectra acquired
additionally do not result in correct or self-consistent CPD values~\cite{Koenig09, SOM}.

As TIBB($V$) is a monotonic function of the
applied sample bias, shifted with respect to zero bias by the CPD, we can
attribute a negative (positive) TIBB($V$) to any
sample bias below (above) $+0.64$~V, and TIBB$(0.64~\text{V})=0$ for tip apex \#1.
The voltage dependence of TIBB($V$) is shown in Fig.~1(b), where
we used $V_\text{CPD}$
as an input parameter to a one-dimensional
Poisson-equation solver developed by Feenstra~\cite{semitip}.
The flat-band voltage [and the corresponding zero crossing of TIBB$(V)$] is indicated
by a vertical (horizontal) dash-dotted line.
We point out that the magnitude of TIBB($V$) is still uncertain, as it depends on the
geometry of the tip, which cannot be easily extracted from KPFS.
%-----------\section{Addressing single Z\lowercase{n} acceptors near the G\lowercase{a}A\lowercase{s}(110) surface}
\begin{figure}[]
\includegraphics[]{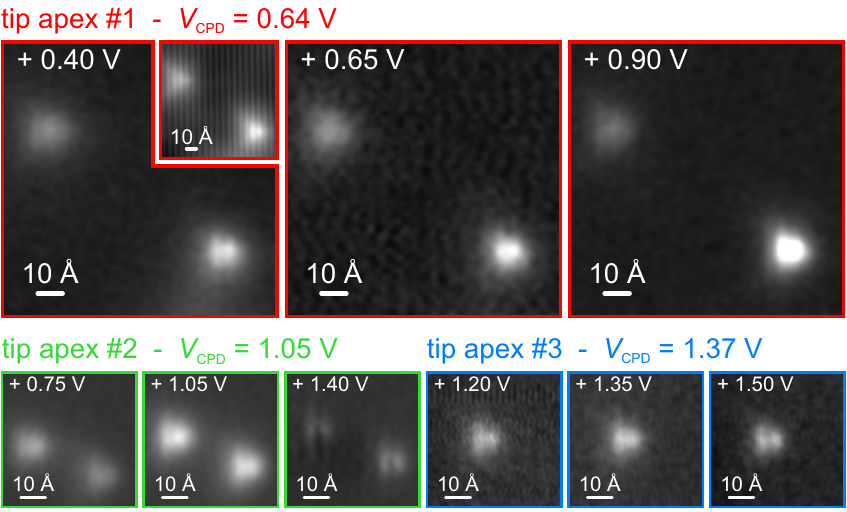}
\caption{
         $dI/dV$ maps of subsurface acceptors acquired with tip apices~\#1 (inset: constant-current topography,
	$V=1.5$~V, $I=20$~pA) to~\#3.
         For each apex, maps are acquired at voltages below, at, and above
	the corresponding flat-band voltage (sample biases as indicated; the gray scale is identical for each apex).
	For all apices, within all band-bending regimes, a similar triangular feature of enhanced conductance is observed.
}
\end{figure}

Now, we investigate the electronic transport through Zn acceptors buried below the GaAs(110) surface
by means of STS in view of the
experimentally determined CPD for tip apex~\#1.
Figures~1(c) and 1(d) show (spatially averaged) $I(V)$ and differential conductance [$dI/dV(V)$] spectra,
recorded away from and atop a subsurface acceptor, as is indicated by the colored
lines in the constant-current topography shown in the inset.
Both spectra
show {\it p}-type semiconducting characteristics, with a valence-band-related
current onset below 0~V and a conduction-band-related current onset above 1.5~V.
In contrast to the spectra as acquired away from the acceptor,
the spectra acquired atop the Zn acceptor also
show nonzero current and conductance within a large bias interval located within
the semiconducting band gap.
Whereas this has been reported before~\cite{Mahieu05, Loth06, Wijnheijmer10},
the nonzero conductance has so far never been related to quantitative contact potential difference measurements.
For this particular tip apex the flat-band condition has been unambiguously determined to be +0.64~V;
therefore we immediately see that acceptor-related conductance is present for negative, zero, and positive TIBB($V$).
The same holds true for data acquired with tip apices \#2 and \#3
which show distinctly different values of the CPD (1.05 and 1.37~eV, respectively); see Fig.~1.

To map out the spatial dependence of the acceptor-related enhanced conductance,
we have recorded differential conductance maps.
In Fig.~2, we show $dI/dV$ maps acquired with tip apices~\#1 to~\#3, recorded at
bias voltages well below, right at, and well above the corresponding
flat-band voltage.
In accordance with previous experiments, we observe a triangular feature
of enhanced conductance at the position of the dopant atom~\cite{Mahieu05}.
Most notably, for all apices, a similar pattern of enhanced conductance is observed for negative, zero and positive band bending.
The similarity present in different band-bending regimes suggests that
{\it one} conduction mechanism is responsible for all of them.
Most importantly, the polarity of the sample bias of the differential conductance maps
for all three band-bending regimes remains the same,
such that the occurrence of the same conduction mechanism is not related to bipolar tunneling~\cite{Wu04, Mahieu05, Morgenstern08}.
The basis of most pictures used so far in this context
is a single {\it isolated} acceptor level being shifted by TIBB($V$) against the Fermi level.
In this picture, the occupation of the level has to change when shifted across the Fermi level,
which determines whether or not a particular channel can contribute to transport,
independent of the further details of the model~\cite{Loth06, Wijnheijmer10}.
Hence, in these pictures, no transport mechanisms can be active in {\it all} three regimes
for one particular sample bias polarity.
However, our conductance spectra, related to the flat-band voltage, in combination with the $dI/dV$ maps,
indeed suggest that one conduction mechanism is active in all three regimes.

%-----------\section{Disuccsion & Theory}
To resolve this controversy, we treat the system as
a linear chain of $N$ equidistant acceptor states between the microscope's tip and the bulk of the sample; see the inset of Fig.~3(a).
In this picture,
three energies are important for the description.
(i) The band bending shifts the on-site energy $\epsilon_i$ of each acceptor state, depending on its position below the surface.
This shift is zero deep inside the bulk and is assumed to increase quadratically towards the surface,
where it reaches TIBB($V$) as plotted in Fig.~1(b).
(ii) Adjacent acceptor states are coupled via a hopping parameter $t$, which we have taken,
in accordance with the impurity band width of about 20~meV, to be $t=5$~meV.
(iii) In our system, the on-site Coulomb energy $U$ of an
isolated acceptor is estimated to be on the order of 10~meV, given a size of the acceptor state of about 20~\AA\,
and a dielectric constant of GaAs of about 13~\cite{bookAdachi, bookKamimura, bookMott}.
The Hamiltonian in the
Anderson-Hubbard model reads~\cite{Anderson61, Hubbard63, Arseev01}
\begin{equation}
\begin{split}
H &= \sum_{i = 1}^{N} \sum_{\sigma}\epsilon_i c^\dagger_{i\sigma}c_{i\sigma}
-t\sum_{i=1}^{N-1}\sum_{\sigma} \left(c^\dagger_{i\sigma}c_{i+1\sigma} + c^\dagger_{i+1\sigma}c_{i\sigma}\right)\\
&+ U\sum_{i=1}^{N} \left(c^\dagger_{i\uparrow}c_{i\uparrow}-\tfrac{1}{2}\right)
\left(c^\dagger_{i\downarrow}c_{i\downarrow}-\tfrac{1}{2}\right)
\nonumber
\end{split}
\end{equation}
where $c^\dagger_{i\sigma}$ creates and $c_{i\sigma}$ annihilates an electron of spin $\sigma$ on the $i$th acceptor.
Here, we choose $N=5$~\cite{SOM}.
The rest of the acceptor states and the valence and the conduction bands have been modeled
as an electron bath with the respective densities of states. The metallic tip has been treated analogously,
having a constant density of states.
Further, we assume that the tunneling between tip and acceptors is restricted to the most superficial acceptor
[cf. the inset of Fig.~3(a)].
All foremost acceptors are coupled to the bulk of the sample.
We note that energy dissipation is expected to occur via the inelastic excitation of vibrons~\cite{vibronic}.
The dynamics of the system is understood as a sequence of tunneling events from (to) the tip or the bulk of the sample
which increase (reduce) by 1 the number of electrons populating the foremost acceptors.
The method of choice for the description of these sequential tunneling dynamics is thus the master equation approach~\cite{SOM}.
In accordance with the experimental situation, the tunneling rate $\Gamma^{T}$ to and from tip states
is by far the smallest, and thus the foremost acceptors are essentially in equilibrium with the bulk.
Moreover, for $V \approx V_{\textrm{CPD}}$, electrons cannot tunnel from the
foremost acceptors to the tip
since all transport resonant levels lie far below the tip's Fermi level.
Under these assumptions, the current through the system takes the form
$I = e\Gamma^{T}\left(2\,-\langle n_{N}\rangle\right)$, where $\langle n_{N}\rangle$
is the average occupation of the most superficial acceptor [see Fig.~3(b)].
\begin{figure}[]
\includegraphics[]{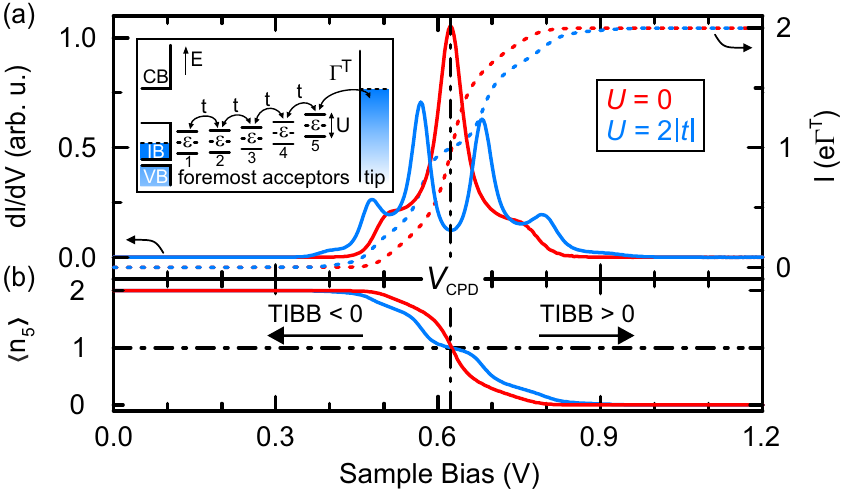}
\caption{\label{fig3}
	Simulated electron transport within the junction.
	(a) Calculated $I(V)$ (dashed lines) and $dI/dV(V)$ (solid lines) spectra for vanishing ($U=0$) and nonvanishing
	on-site Coulomb energy ($U=2|t|$).
	The simulation yields nonzero current and conductance in a broad voltage range including
	negative, zero, and positive band bending. The inset sketches the relevant energies of the single-particle
	levels used as input for the many-body calculations.
	(b) Calculated average population $\langle n_5 \rangle$ of the foremost acceptor.
}
\end{figure}

This many-body approach ensures that the
{\it gradual change} of (i) the effective electronic coupling,
(ii) the localization, and (iii) the charging energy
of the relevant states
as a function of bias voltage is inherently captured.
Figure~3(a) shows the simulated $I(V)$ and $dI/dV(V)$
spectra for different values of $U$.
Neglecting charging energy ($U=0$),
it shows enhanced current and conductance
within a large bias range in different band-bending regimes, for one sample bias polarity only, which is
in accordance with our experimental findings.
This phenomenon can be understood as follows.~%
For $V \simeq V_\textrm{CPD}$,
TIBB($V$) is smaller than the impurity-band width,
and hence the foremost acceptor state is still part of the impurity band,
even if slightly detuned from the bulk impurity states.
In this voltage region, the Fermi level still remains inside this
impurity band that extends to the foremost acceptor, and hence, finite conductance is observed.
The size of this bias voltage range around $V_\textrm{CPD}$ is given by the
impurity-band width $4t$ divided by the lever arm $\alpha=$~TIBB$(V)/V$.
For $U\neq 0$,
several peaks and dips appear in the spectra
as opposed to just a single broad peak that is observed for $U=0$.
Close to $V_{\textrm{CPD}}$,
the single occupation of the foremost acceptor
prevails until $|$TIBB$(V)$$|$ overcomes $U/2$.
Hence, the average population and the current develop a plateau around $V_{\textrm{CPD}}$ of width $U$ divided by $\alpha$.
Slight modifications in the on-site energies $\epsilon_i$ and in the tunneling coupling $t_{ij}$ between
adjacent states $i$ and $j$
result in variations of the relative peak heights as well as their
positions with respect to $V_\textrm{CPD}$~\cite{SOM}.

The simulated spectra in Fig.~3(a) are
in qualitative agreement with our experimental ones~\cite{tip_prep},
showing enhanced conductance in all three band-bending regimes.
The experimental spectra show enhanced conductance over an even wider bias range than our theory predicts.
Whereas $U$ and $\alpha$ may differ from the values anticipated here,
we note that electron-vibration coupling~\cite{vibronic} could also play an important role,
the incorporation of which goes beyond the scope of our model.

Finally, we note that
the knowledge of the CPD in our experiments also sheds new light onto the interpretation
of the observation of charge density oscillations around acceptors~\cite{deKort01}, as is discussed
in the Supplemental Material~\cite{SOM}.

%-------------------\section{Summary}
In summary, the use of combined STS and KPFS allows us to unambiguously relate
the conductance properties of shallow acceptors buried in GaAs
to the energy scale of the system,
by measuring the flat-band voltage.
These measurements show that the
voltage range of enhanced acceptor-induced conductance
spans three different band-bending regimes,
ruling out previous conceptions of electronic transport
used in this context~\cite{Mahieu05, Loth06, Wijnheijmer10}.
This experimental finding
requires a theoretical description which inherently captures the crossover between the acceptor
state being part of an impurity band for zero band bending, and the acceptor state being split off and localized for strong negative
or positive band bending, respectively.
Transport calculations based on an Anderson-Hubbard model yield spectra in qualitative agreement with our experiments.
We expect that this combination of Kelvin probe and scanning tunneling spectroscopy can
shed new light on the energetics in cross-sectional
STM experiments far beyond the specific model system studied here.
%-------------------\section{Acknowledgment}

The authors thank S. Rolf-Pissarczyk, D.~Gohlke, F.~J.~Gie\ss ibl, and M. Morgenstern
for valuable discussions and M.~Utz, M.~Neu, and A.~P\"{o}llmann
for instrumentation
support. Funding from the Volkswagen Foundation
(Lichtenberg Program) and the Deutsche Forschungsgemeinschaft
(SFB~$689$, SPP~$1243$, and SPP~$1285$)
is gratefully acknowledged.

%-------------------------------REFERENCES----------------------------------%
\bibliographystyle{apsrev4-1}
%\bibliography{)}

\newpage

\preprint{}
\begin{center}
\large{SUPPLEMENTAL MATERIAL}
\end{center}

%---------------------------\section{Introduction}

%------------------------------------------\section{Experimental}

In this supplemental material, we present details on the experimental setup and
on the determination of the uncertainty of the KPFS measurement.
A comparison of CPD values as inferred from $I(z)$-spectroscopy and as determined from KPFS is provided.
Further, we present data on the bias dependence of the onset of local density of states oscillations around acceptors.
Finally, a detailed description of the Anderson-Hubbard model is presented.

\subsection{Experimental Setup}

In our setup, the bias voltage $V$ is applied to the sample with respect to the microscope's tip.
The contact resistance between the GaAs sample and the sample-holder was found to be in the k$\Omega$ range,
at a temperature of 5~K, which is many orders of magnitude below the resistance of the tunnel junction.

To record all current $I$ versus $V$ spectra and differential conductance (d$I$/d$V$)
maps at the same absolute tip height
above the GaAs(110) surface, we first opened the feedback loop of the STM at  $V=1.8$~V, $I=20$~pA,
with tip located away from acceptors, and then moved the  tip to the position where spectra or d$I$/d$V$ maps were taken.
For d$I/$d$V$ maps acquired with tip apex~\#2,
the tip-sample distance was decreased by $\Delta z = 0.25$~\AA\ once the feedback-loop of the STM was interrupted.
$I(V)$ spectra are (except where stated otherwise) averaged over each 10 single spectra,
acquired along a line in a specific sample region.
d$I$/d$V(V)$ spectra are numerically derived from the $I(V)$ data, data points
are averaged over a bias range of 50 mV.
For d$I$/d$V$ maps, we used lock-in technique with 50~mV peak-to-peak at 166~Hz.

For KPFS, the tip was retracted by $\Delta z$~=~5~\AA\, once the feedback-loop of
the STM was interrupted at $V=1.8$~V, $I=20$~pA.
For such increased distance, the CPD measurement
is not influenced by atomic-scale variations of the sample surface, but depends
only on the long-range electrostatic interaction between tip and sample~\cite{Gross09}.
Moreover, at such increased distance, the tunneling current is zero,
such that we can exclude any influence of the current on the $\Delta f(V)$ signal~\cite{Weymouth11}.
As oscillation amplitudes $(A)$ of the qPlus force sencor (spring constant $(k) \approx 1.8\times 10^3$~N/m$^{-1}$,
resonance frequency $(f_0)\approx26$~kHz, %$(f_0)=26031$~Hz,
quality factor $(Q) \approx10^4$), we used $1.2$~\AA, $2.5$~\AA, and $0.5$~\AA\,
for measurements with tip apices \#1, \#2, and \#3, respectively.

The apex of the microscope's tip was treated by controlled indentations into the (111) surface of a Cu single crystal.
The Cu single crystal, mounted on a dual sample-holder next to the GaAs sample, was prepared by repeated
sputter and annealing cycles, which were performed prior to the GaAs cleavage.

To ensure that the apex of the microscope's tip used to probe the GaAs(110) surface is
identical to that for the Cu(111) surface, the relative alignment between both planar
surfaces has been adjusted to be accurate within $3^\circ$.
To test whether the tip apex was indeed the same when probing both surfaces,
we transferred a CO molecule to the apex on one surface and
moved the tip over to the other sample surface.
There we  observed the high lateral resolution that can be attributed to a CO-functionalized tip.

\subsection{Kelvin probe force spectroscopy for different tip-sample distances}
\begin{figure}[b]
\includegraphics[]{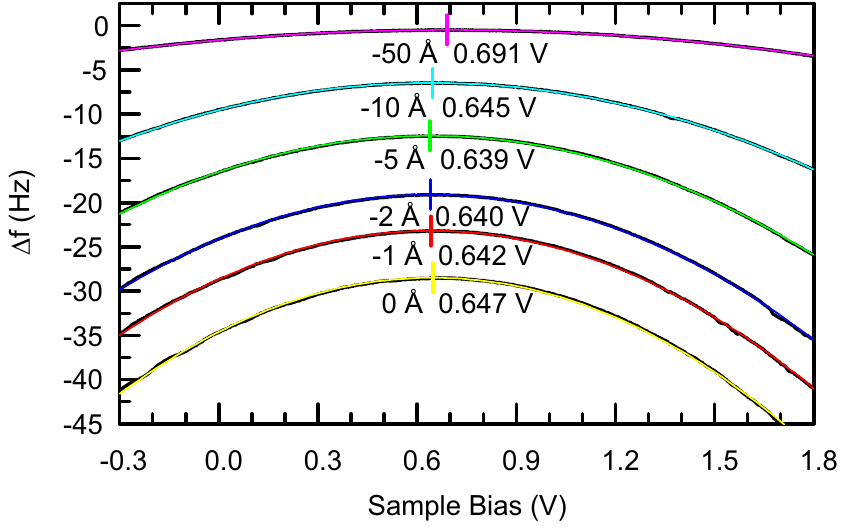}
\caption{\label{fig1}
           KPFS for different tip-sample distances.
           The frequency shifts $\Delta f(V)$ (black) are fitted by
           Kelvin parabolas (colored). The standard deviation of each fit is below 1~mV.
           The relative tip-sample distances $\Delta z$
           and the corresponding contact potential differences $V_{\textrm{CPD}}$
           are indicated below each curve.
           The total variation of $V_{\textrm{CPD}}$ is very small compared to
           sample bias range considered in our experiment.
}
\end{figure}

KPFS measurements can be subject to averaging effects,
if areas of different work function contribute to the measurement~\cite{Sadewasser}.
As our sample is atomically flat and homogeneous, such artifacts can
arise only if the work function of the tip is spatially inhomogeneous.
If this is the case, the measured contact potential difference (CPD) is a weighted
average over different tip regions of different work functions.
As in KPFS the ratio of contribution of different parts of the tip strongly depends
on their relative distance to the sample, such  inhomogeneity would result
in different values for the CPD measured for different absolute tip-sample
distances~\cite{Sadewasser}.
Figure~S\ref{fig1} shows KPFS measurements for different absolute tip-sample distances acquired with tip apex \#1.
At a setpoint of $V = 1.8$~V and $I = 20$~pA the feedback-loop of the STM
was opened. The tip was then retracted by different values of $\Delta z$ ranging
from $0$ to $50$~\AA, and $\Delta f(V)$ was recorded.
We find  the peak positions of the Kelvin parabolas to differ by less than 45~mV in total.
For tip apices  \#2 and \#3 we proceeded likewise, finding total variations of below 23~mV and 52~mV,
respectively.
This variation of the CPD for individual apices is very small compared to the bias voltage range considered here.
Accordingly, for the tip apices used, inhomogeneities can be largely ruled out, and we expect the error of
$V_{\textrm{CPD}}$ from this side
to be on the order of a few 10~meV.

\subsection{Determination of the absolute error of the KPFS measurements}
\begin{figure}[b]
\includegraphics[]{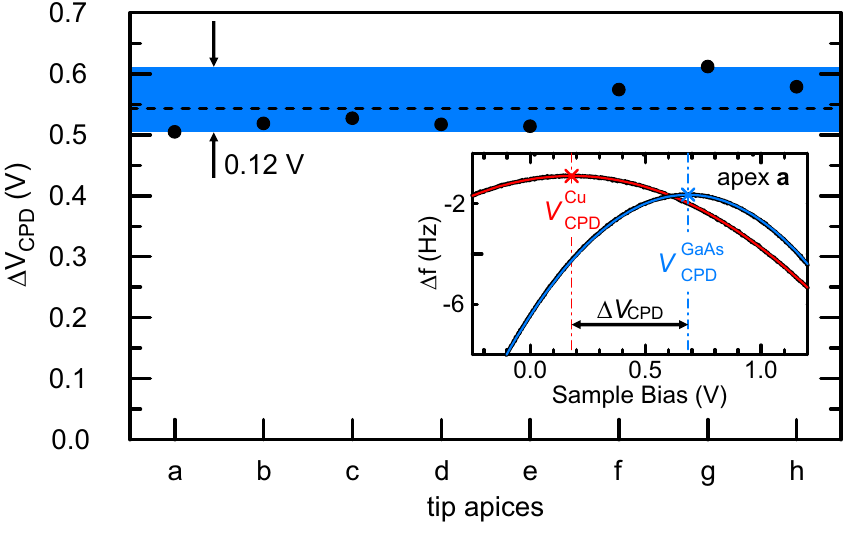}
\caption{\label{fig2}
	Difference of the contact potential difference acquired on Cu(111) and on GaAs(110),  $\Delta V_{\rm CPD}$,
	for eight individual tip apices. The horizontal dashed line indicates the arithmetic average of $\Delta V_{\rm CPD}$ (0.54~V),
	its absolute variation is found to be 0.12~V.
         The inset shows, as an example, $\Delta f(V)$ data acquired with tip apex~{\bf a} (black lines),
	parabolic fits are indicated (colored).
}
\end{figure}
To determine the absolute experimental uncertainty of our KPFS measurements on GaAs(110),
we used KPFS data acquired on the clean Cu(111) surface (used for tip preparation) as a reference.
For an individual tip apex, we determined the voltage corresponding to the
CPD on both surfaces, {\it i.e.} $V_{\rm CPD}^{\rm GaAs}$ and $V_{\rm CPD}^{\rm Cu}$~\cite{parameters, error}.
Figure~S\ref{fig2} shows, for eight individual tip apices (with tip work functions varying by 0.4~eV),
$\Delta V_{\rm CPD} = V_{\rm CPD}^{\rm GaAs} - V_{\rm CPD}^{\rm Cu}$.
In total, $\Delta V_{\rm CPD}$ is found to vary by $0.12$~V, at an arithmetic averaged value of $0.54$~V.
From the literature, $\Delta V_{\rm CPD}$
is expected to be
\begin{equation}
\begin{split}
\Delta V_{\rm CPD} &=  \frac{1}{e}\left[\chi_{\textrm{GaAs(110)}} +  (E_{\textrm C} - E_{\textrm F})-\phi_{\textrm{Cu(111)}}\right]\\
&=4.07{~\rm V}+1.49{~\rm V}-4.94{~\rm V}\\
&=0.62{~\rm V}
\end{split}
\end{equation}
with $\chi_{\textrm{GaAs(110)}}$ and $(E_{\textrm C} - E_{\textrm F}) $ being the electron affinity
of GaAs(110) and the difference between conduction band edge and Fermi-level in Zn-doped GaAs, respectively,
and  $\phi_{\textrm{Cu(111)}}$ being the work function of the clean Cu(111) surface.
From the literature, we have taken 4.94~eV for the work function of Cu(111)~\cite{BookLinde},
4.07~eV for the electron affinity of GaAs(110)~\cite{BookAdachi}, and $(1.52-0.03) $~eV~=~1.49~eV for the
energetic difference between conduction band edge and Fermi-level in Zn-doped GaAs, with the Fermi-level
located 0.03~eV above the edge of the valence band~~\cite{BookAdachi, bookSchubert}.

Accordingly, we estimate the absolute error of the voltage corresponding to the flat-band condition, $V_{\rm CPD}$,
as extracted from KPFS on GaAs(110), to be below~$0.12$~V, which is small compared to the voltage scale of interest.

\subsection{Comparison of $I(z)$-spectroscopy and KPFS}
\begin{figure}[b]
\includegraphics[]{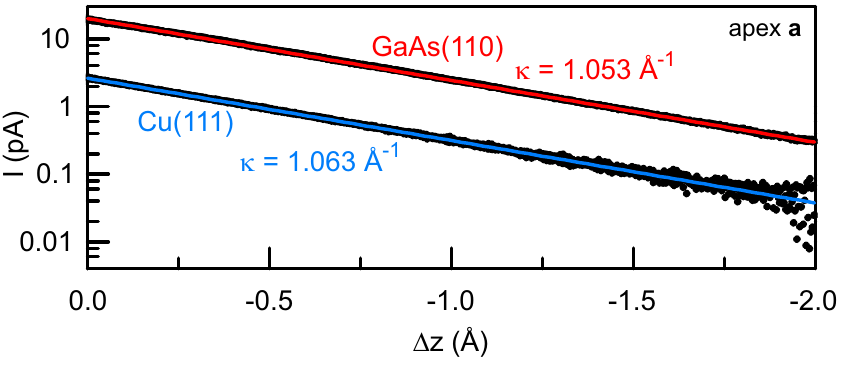}
\caption{\label{fig7}
         $I(z)$-spectroscopy on Cu(111) and on GaAs(110) surfaces (black, semilogarithmic plot),
	single-exponential fits and corresponding   		
	values of the inverse decay length
	$\kappa$ are indicated (color).
}
\end{figure}

In the literature, in the context of acceptor-induced enhanced conductance,
the dependence of the tunneling-current $I$ on tip-sample distance $z$ was used to infer the CPD~\cite{Loth06, Loth07, LothPhD, Wijnheijmer10}.
%\newpage
Figure S\ref{fig7} shows a semi-logarithmic plot of $I(z)$-spectra,
acquired with tip apex {\bf a} (see inset of Fig.~S\ref{fig2} for corresponding $\Delta f(V)$ data) on Cu(111) and on GaAs(110).
On Cu(111), the feedback loop of the STM was interrupted at a current of 2.5~pA and a sample bias of $+50$~mV, while
%For KPFS on Cu(111), the tip was retracted by 5~\AA once the feedback loop of the STM was interrupted at a
%current of 2.5~pA and a sample bias of 50~mV.
%For $I(z)$
on GaAs(110) the feedback loop was interrupted at a current of 20~pA and a sample bias of $+1.6$~V.

In $I(z)$-spectroscopy, the tunneling current is assumed to
decay exponentially with increasing tip-sample distance, $I\propto \exp(-2\kappa z)$.
The inverse decay length $\kappa$ is linked to the
apparent barrier height $\overline\phi$ by the expression
$\kappa=\sqrt{2m\overline\phi/\hbar^2}$,
where $m$ is the mass of the free electron.

Whereas comparative $I(z)$-spectroscopy of different sample areas can
provide a qualitative measure for the changes in work function~\cite{Olsson05}, %, Vitali Nature Materials 9, 320 (2010)},
some caution is in order when trying to quantitatively interpret the inverse decay length $\kappa$~\cite{Koenig09}.
On semiconductors, things are expected to be even more complicated, as
TIBB$(V)$ changes also with distance $z$, which can additionally affect the $I(z)$-spectra, by changing the number of states
in the sample available for tunneling.
The derivation of an actual value for the CPD from $I(z)$-spectra
depends strongly on the model for tunneling that is applied~\cite{Loth06, Loth07, LothPhD, Wijnheijmer10}.
For easy comparison,
we here use a model solely based on geometrical considerations, which was
applied previously by Loth {\it et al.}~\cite{Loth06, Loth07, LothPhD}.

%For KPFS on GaAs(110), the tip was retracted by 5~\AA once the feedback loop of the STM was interrupted at a
%current of 20~pA and a sample bias of 1.6~V.
%Assuming an exponential decay of the current $I$ with increasing
%tip-sample distance $z$, $I\propto \exp(-2\kappa z)$, we extract, from a single-exponential fit,
%the inverse decay length $\kappa=\sqrt{2m\overline\phi/\hbar^2}$, where $m$ is the mass of the free electron and
%$\overline\phi$ is the apparent barrier height.

On GaAs(110), % following the method proposed by Loth {\it et~al.}~\cite{Loth07}
for $ eV> (E_{\textrm{C}}-E_{\textrm{F}})$,
the apparent barrier height is connected
with the tip's work function, $\phi_{\textrm{tip}}$, via~\cite{LothPhD}:
\begin{equation}
\overline\phi =  \frac{\chi -[eV - (E_{\textrm{C}}-E_{\textrm{F}})]+\textrm{TIBB}(V)+\phi_{\textrm{tip}}}{2}
\label{eqTIBB}
\end{equation}
%where TIBB($V$) is the tip induced band bending as introduced in the main text.
In this equation, TIBB($V$) gives rise to a small correction (below 0.1~eV, c.f.~Fig.~1(b), positive sample bias), which
is unimportant for our comparison and which we therefore neglect.
%As all our $I(z)$-spectra are taken at positive sample biases, exceeding the band gap of GaAs,
%we expect TIBB($V$) to be only a minor correction to equation~\ref{eqTIBB}.
%Indeed, from the TIBB($V$) calculations shown in Fig.~1(b) (main text),
%we see that in this bias range TIBB($V$) is typically below 100~meV.
%That given, we, for simplicity, neglect the TIBB($V$) term in equation~\ref{eqTIBB}.
Solving equation \ref{eqTIBB} for $\phi_{\textrm{tip}}$
one can then calculate $V_{\rm CPD}^{\rm GaAs(110)}$ via:
\begin{equation}
V_{\rm CPD}^{\rm GaAs(110)} = \frac{1}{e}\left[\chi + (E_{\textrm{C}}-E_{\textrm{F}}) - {\phi_{\textrm{tip}}}\right]
\end{equation}

On Cu(111), we proceed likewise,
%in the model of a trapezoidal barrier, using the Wentzel-Kramers-Brillouin approximation,
connecting the apparent barrier height $\overline\phi$
with the work functions of tip and sample via:
\begin{equation}
\overline\phi = \frac{1}{2}\left(\phi_{\textrm{tip}}+\phi_{\textrm{Cu(111)}}\right)
\end{equation}
leading to
%On Cu(111), using the Wentzel-Kramers-Brillouin approximation, $V_{\rm CPD}^{\rm Cu(111)}$ is determined via:
\begin{equation}
V_{\rm CPD}^{\rm Cu(111)} = \frac{2}{e}\left(\phi_{\textrm{Cu(111)}}-\overline\phi \right)
\end{equation}

In Table~\ref{table1}, we present, for tip apex~{\bf a}, $V_{\rm CPD}$ as determined from KPFS and as inferred from the $I(z)$-spectra
shown in Fig.~S\ref{fig7}.

\begin{table}[h!]
\begin{center}
\begin{tabular}{l c c}
 &\parbox{1.8 cm}{\centering KPFS} & \parbox{1.8 cm}{\centering $I(z)$}\\
\hline
$V_{\rm CPD}^{\rm Cu(111)}$ (V)\;\; & $0.18\pm0.12$ & 1.27 \\
\hline
$V_{\rm CPD}^{\rm GaAs(110)}$ (V)\;\; & $0.68\pm0.12$ & 1.06 \\
\end{tabular}
\end{center}
\caption{\label{table1}Comparison of $V_{\rm CPD}$, for one individual tip apex,
as determined from KPFS and inferred from $I(z)$, on Cu(111) and on GaAs(110).}
\label{table1}
\end{table}

In addition, for tip apex~\#3 (main text), we record a series of 20 $I(z)$-spectra on GaAs(110),
equidistant spaced along a $40$~\AA\, long line, oriented parallel to the crystallographic [001] direction,
located away from dopant atoms or defects.
Using the model described above,  we inferred $V_{\rm CPD}$ from each individual spectra, and subsequently
calculate mean and standard deviation of the series of spectra.
%for each individual $I(z)$-spectra, we inferred $V_{\rm CPD}$.
% of 20 spectra.
This procedure was repeated three times, varying
the sample bias at which the feedback-loop of the STM was interrupted at a current of 20~pA.
%before the tip-sample distance was changed.
The corresponding results are presented in Table~\ref{table2}.

\begin{table}[h!]
\begin{center}
\begin{tabular}{ l c c c c }
 &{\multirow{2}{*}{\parbox{2.2 cm}{\centering KPFS}} }&\multicolumn{3}{c}{$I(z)$} \\
\cline{3-5}
sample bias (V)  & &\parbox{2.2 cm}{\centering $+1.5$} &\parbox{2.2 cm}{\centering $+1.6$}&\parbox{2.2 cm}{\centering $+2.0$} \\
\hline
$V_{\rm CPD}^{\rm GaAs(110)}$ (V)  & $1.37\pm0.12$ & $0.65\pm0.43$ & $-0.01\pm0.30$ & $1.81\pm0.14$\\
\end{tabular}
\caption{\label{table2}Comparison of $V_{\rm CPD}$, as determined from KPFS and inferred from $I(z)$-spectroscopy for tip apex~\#3.
$I(z)$-spectroscopy was performed for three different setpoint-values of the sample bias.}
\end{center}
\end{table}

The above analysis shows that the CPD values derived from $I(z)$-spectroscopy
are neither consistent with the differences in work function for Cu(111) and GaAs(110)
as known from the literature
nor are self-consistent, when comparing data acquired at different sample biases.
We note that the difference in CPD between Cu(111) and GaAs(110)
as derived from $I(z)$-spectroscopy
even has the wrong sign.

Finally, we note that including mirror charge effects to the model
of tunneling, as proposed by Wijnheijmer {\it et al.}, is expected to considerably lower the
flat-band voltage $V_{\textrm{CPD}}$ (by~$\sim 1$~eV)~\cite{Wijnheijmer10},
which again would not result in correct values for $V_{\textrm{CPD}}$.

\subsection{Spatially resolved d$I/$d$V(V)$-spectra}

\begin{figure}[]
\includegraphics[]{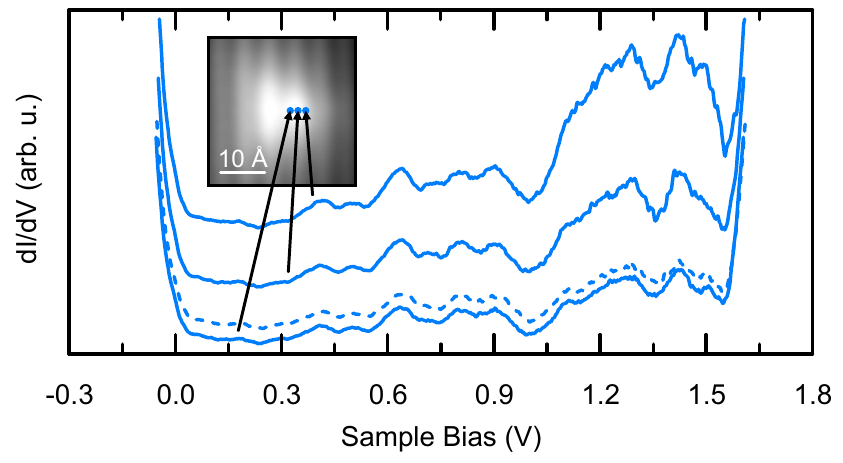}
\caption{\label{fig4}
Spatially resolved d$I/$d$V(V)$ spectra acquired atop a subsurface acceptor (inset: $V = 1.5$~V, $I = 20$~pA).
Three spectra (solid lines) were acquired at three different sample positions very close to one another.
The spectra are vertically offset for clarity. At the same position of the lowest of these spectra,
an additional spectrum (dashed line) was acquired.
}
\end{figure}

In the main text, in Fig.~1(c) we show (numerically derived)
d$I/$d$V(V)$ spectra, spatially averaged over 10 single spectra, acquired along
a line at a specific sample region.
To prove the reproducibility of our data,
in Fig.~S\ref{fig4} we show four individual, spatially resolved (numerically derived)
d$I/$d$V(V)$ spectra (vertically offset for clarity), acquired with tip apex~\#1.
%The spectra are which were acquired with tip apex~\#~1.
The two lowest (dashed and solid) were acquired at the same %sample
position, whereas the others were taken
after slightly moving the tip laterally as indicated in the inset.
The spectra acquired at the same position are almost identical, the others continuously evolve
while moving the tip laterally.
Whereas this clearly demonstrates the high reproducibility of our data,
it is still likely that some part of the fine structure of the spectra is related to
a non-constant tip density of states.
To minimize the latter influence in our spectra,
we used a dual sample holder in our experiments, which allowed us to
prepare a metal-terminated tip on Cu(111) before moving to the semiconductor and taking spectra there.

%\begin{figure}[]
%\includegraphics[]{figuresSOM/fig10SOM.pdf}
%\caption{(color online) \label{fig10}
%	Contact potential difference and acceptor induced conductance.
%	(a) Frequency shift measured as a function of sample bias, $\Delta f(V)$ (black lines),
%	a parabolic fit and the corresponding flat-band voltage $V_{\textrm{CPD}}$ are indicated (blue).
%	For any sample bias below (above) $V_{\textrm{CPD}}$, the tip induced band bending is negative (positive).
%	(b) and (c) show $I(V)$ and d$I/$d$V(V)$ spectra away from (dashed) and atop a sub-surface acceptor (solid),
%	positions are indicated in the constant-current STM image (inset: $U= 1.51$~V, $I=20$~pA).
%	Acceptor-induced enhanced current and conductance is observed in negative, zero and positive band bending regimes.
%}
%\end{figure}

%\subsection{Complete set of conductance maps}

%\newpage
%Figure~S\ref{fig4} shows bias-dependent constant-height conductance maps acquired with tip apices \#2 and \#3.
%Similar to what is observed with tip apex \#1, a similar feature of enhanced conductance is present in multiple band bending regimes.
%\newpage
%\begin{figure}
%\includegraphics[]{figuresSOM/fig4SOM}
%\caption{\label{fig4}
%           d$I$/d$V$ maps acquired with tip apices \#2 and \#3
%           (corresponding flat-band voltages and sample biases are indicated).
%          As with tip apex \#1, a similar feature of enhanced conductance is observed for bias voltages below, at, and above
%         the flat band voltage.
%}
%\end{figure}

\subsection{Local density of states oscillations}
\begin{figure}[h!]
\begin{center}
\includegraphics[]{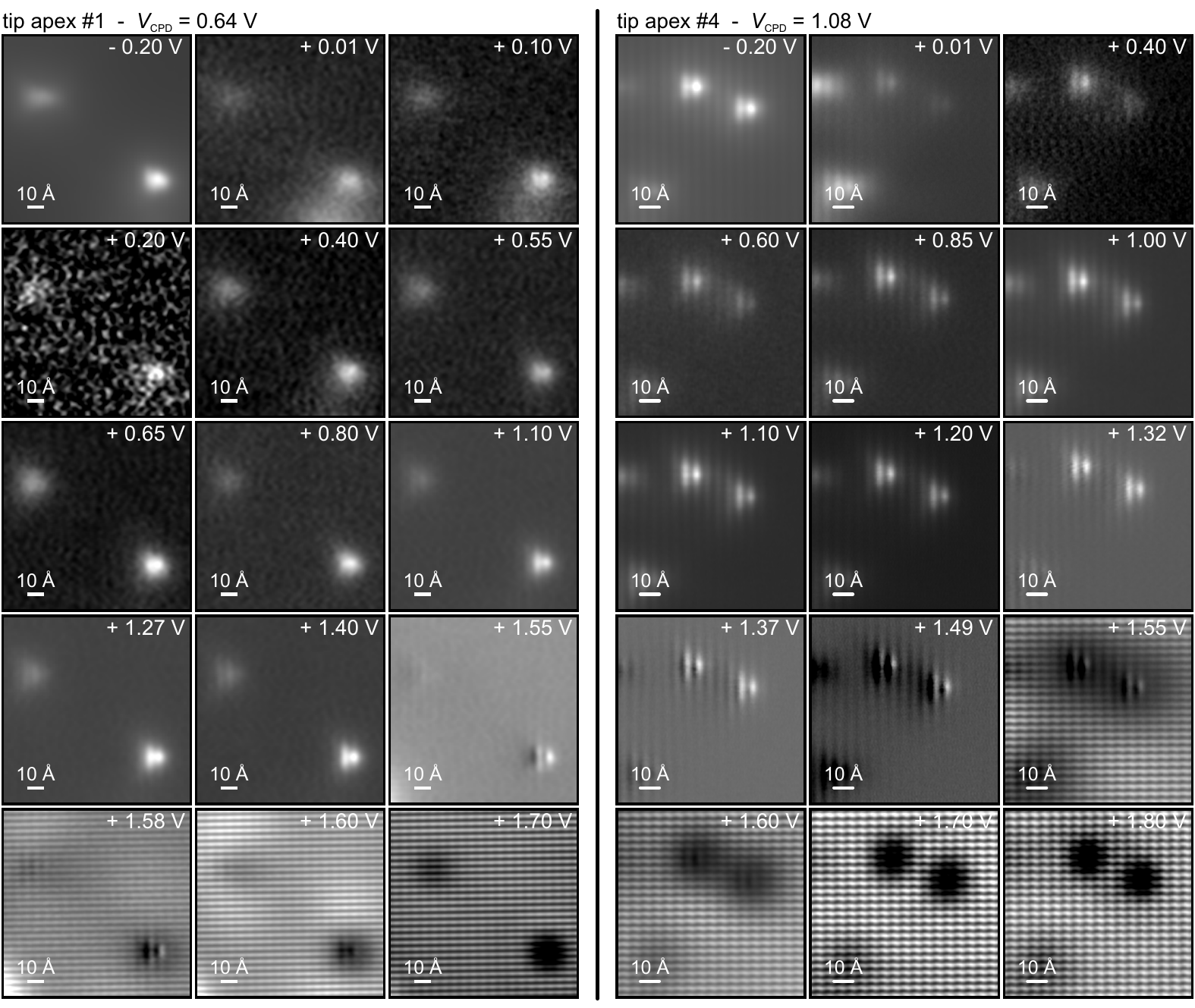}
\caption{\label{fig3}
         d$I$/d$V$ maps (sample biases as indicated),
	acquired with two different tip apices~\#1 (left) and \#4 (right),
	with markedly different flat-band voltages of 0.64~eV and 1.08~eV, respectively.
         For both apices, within a large sample bias window, including positive sample biases
	below, at, and above the flat-band voltage, a triangular feature of enhanced conductance is observed at
	the position of the dopant atoms.
	Local density of states oscillations are, for both apices, observed only at sample biases corresponding to
	tunneling of electrons from the tip to the conduction band of the sample ({\it i.\,e.} for $V > 1.5$~V).
	%The dark depressions observed at $+1.55$~V corresponds to negative differential conductance.
	%In a large sample bias window, including biases
         %below, at, and above the flat-band voltage of $V_{\textrm{CPD}}=0.64$~V,
         %a triangular feature of enhanced conductance is observed at the position of the
         %dopant atoms.
}
\end{center}
\end{figure}
Figure~S\ref{fig3} shows two sets of bias-dependent d$I/$d$V$ maps acquired with two different tip apices,
namely tip apex \#1 and an additional tip apex, labeled \#4~\cite{paramtip4}.
The images acquired with tip apex~\#1 are recorded on the same area as those shown in Fig.~2,
main text.

First, we notice that for both tip apices, which have distinctly different flat-band voltages of
0.64~V and 1.08~V, respectively, we observe within a large sample bias region (at positive
sample bias polarity), for negative, zero and positive band bending
a similar feature of enhanced conductance at the position of the dopant atoms.
Hence, as already stated in the main text, the observation of enhanced conductance is independent of the sign of TIBB($V$).
In fact, we note that for both tip apices,
the large sample bias region spans almost the entire semiconductor's band gap region ($0 < V < +1.52$~V).

Second, we note that for both apices spatial local density of state oscillations are observed {\it only} for sample biases
for which tunneling of electrons from the tip into the conduction band of the sample is possible, {\it i.\,e.} for $V>+1.5$~V.
Indeed, for both tip apices those oscillations basically occur at the same sample bias as the checker-board-like corrugation
known to be related to tunneling into the conduction band ({\it i.\,e.}, the C$_3$ surface resonance)~\cite{Ebert96}.
%The images are recorded at voltages ranging from $-0.2$~V to $+1.7$~V.
%For this particular tip apex, the flat-band voltage is $+0.64$~V (see Fig.~1 and Fig.~S\ref{fig1}).
%Accordingly, TIBB$(V)$ is negative (zero, positive)
%for any sample bias below (at, above) $+0.64$~V.
%As stated in the main text, within a large sample bias region,
%a similar feature of enhanced conductance is observed at the position of the dopant atoms.
%Hence the observation of enhanced conductance is independent of the sign of TIBB($V$).
%In fact, the large sample bias region spans almost the entire semiconductor's band gap region ($0 < V < +1.52$~V).
In the literature, those oscillations are referred to as charge density oscillations (CDO)~\cite{Wielen96, deKort01, Loth07},
and are expected to occur as soon as holes accumulate below the microscope's tip, {\it i.\,e.} for $V \ge V_{\textrm {CPD}}$.
Those CDO are then explained by the accumulated holes scattering at the core-potentials of the dopant atoms.

From our experiments however, relating the onset-voltage of oscillations in the
local density of states to a quantitative value of the flat-band voltage,
we conclude: whereas the existence of hole accumulation ($V \ge V_{\textrm {CPD}}$) may play a role,
the possibility of tunneling into the conduction band ($V>+1.5$~V) apparently is a necessary requirement for the
observation of local density of states oscillations around acceptors in {\it p}-type GaAs.
We note that our observations are consistent with experimental data found in the literature~\cite{deKort01, Loth07}.

Further, we note that, when comparing these two apices \#1 and \#4 in more detail, one realizes that for
 tip \#1 the occurrence of local density of states oscillations
occurs at slightly higher sample biases as compared to tip \#4 ($\Delta V\simeq 30$~mV).
This slight shift is consistent with the CPD values extracted from KPFS
under the assumption that oscillations are related to tunneling into the conduction band, but inconsistent
under the assumption of being only related to the situation of hole accumulation.
%It has been argued in the literature that
%charge density oscillations (CDO) around acceptors should be observed
%as soon as holes accumulate below the tip, because of upward band bending (TIBB$>0$) [Loth].
%Figures S5 and S7 show constant height d$I$/d$V$-maps acquired with two tips apices (labeled \#1 and \#4) of
%distinctly different flat-band voltages ($V_{\textrm{CPD}}=0.64$ and 1.08~V, respectively).
%For both apices CDO are observed only for voltages, for which tunneling
%of electrons from the tip into the conduction band is possible,
%{\it i.\,e.} for $V>+1.5$~V.
%For both tips the CDO occur basically at the same sample bias as the
%checker-board-like corrugation
%that is known to
%be due to tunneling into the conduction band.

In summary, this is yet another example that the unknown CPD in previous experiments resulted in misinterpretation
of experimental data~\cite{Loth07}.

\subsection{Detailed description of the Anderson-Hubbard model}

In our theoretical analysis we distinguish between bulk and foremost acceptors, the latter being
the ones closer to the tip and thus more strongly affected by TIBB$(V)$. Our tunneling junction is described by the total Hamiltonian:

\begin{equation}
H = H_{\rm acc} + H_{\rm sub} + H_{\rm tip} + H_{\rm tun}
\end{equation}
where $H_{\rm acc}$ describes the foremost acceptors, $H_{\rm sub}$ the rest of the acceptors and the hosting
semiconductor (the substrate), $H_{\rm tip}$ the tip and finally $H_{\rm tun}$ contains the tunneling coupling
between acceptors, tip, and substrate.

The Hamiltonian for the $N$ foremost acceptors includes the Coulomb
interaction, as already stated in the main text; it reads:

\begin{equation}
\begin{split}
\label{eq:Imp_Ham}
H_{\rm acc} &= \sum_{i = 1}^{N} \sum_{\sigma}\epsilon_i c^\dagger_{i\sigma}c_{i\sigma}
-t\sum_{i=1}^{N-1}\sum_{\sigma} \left(c^\dagger_{i\sigma}c_{i+1\sigma} + c^\dagger_{i+1\sigma}c_{i\sigma}\right)\\
&+ U\sum_{i=1}^{N} \left(c^\dagger_{i\uparrow}c_{i\uparrow}-\tfrac{1}{2}\right)
\left(c^\dagger_{i\downarrow}c_{i\downarrow}-\tfrac{1}{2}\right)
\end{split}
\end{equation}
where $c^\dagger_{i\sigma}$ creates an electron of spin $\sigma$ on the $i$th acceptor state,  $\epsilon_i$ is the on-site energy
and $t$ the hopping parameter. Finally, $U$ is the on-site Coulomb repulsion. For the hopping parameter we have taken the value of
$t = 5$ meV in accordance with the accepted impurity band width of roughly $20$ meV. For the Coulomb repulsion term we have assumed
values in the range $U = 10-20$~meV. These values for the on-site Coulomb repulsion are far smaller than the ones of an atom in
vacuum due to the much larger size of the impurity states ($a_B \approx 20$ \AA) and the dielectric constant of the hosting semiconductor
($\epsilon_r \approx 13$). The on-site energy of the acceptor $\epsilon_i$ is the one that is more directly affected by TIBB$(V)$.
We have modeled it as:
\begin{equation}
\label{eq:epsilon}
\epsilon_i = \mu_0 + \left(\frac{i-1}{N-1}\right)^2 {\rm TIBB}(V)
\end{equation}
where $\mu_0$ is the bulk chemical potential of Zn-doped GaAs. Eq.\eqref{eq:epsilon} is
written assuming a linear chain of equidistant acceptors and a quadratic drop off of the tip induced electrostatic
potential inside the semiconductor. %\cite{Feenstra87}.
The deepest acceptor ($i=1$) considered in the
model is not affected by the tip, while the energy of the most superficial one ($i = N$) is shifted exactly
by TIBB$(V)$, as plotted in Fig.~1(b) of the main text.

In a one-dimensional model, the penetration depth of the electric field into the interior of the GaAs sample is,
for the dopant concentration used, estimated to be about $100$~\AA~\cite{Feenstra03}, whereas
the average distance of nearest neighbors of acceptors is about $45$~\AA. In the experimental (three-dimensional) system, assuming a tip-radius of 15~nm, we expect about $10$~acceptors to be
located within the space-charge region caused by tip.
%\del{more than $100$~acceptors are expected to be located within the space-charge region caused by TIBB.}
In our model, due to limitations of computational time,
we restricted ourself to a one-dimensional, linear chain of $N=5$~acceptors.

Due to the particular form of the Coulomb interaction
term, the Hamiltonian \eqref{eq:Imp_Ham} is particle-hole symmetric if $\epsilon_i = 0,\, \forall i$.
This ensures that, in equilibrium, the number of electrons populating the acceptors is always equal to the number of acceptors,
irrespective of the strength of the interaction term $U$. The constant terms appearing in the interaction and ensuring particle-hole symmetry are also physically justified as the contribution of the positive ions located at the position of the acceptor states. The latter ensure the charge neutrality of the system and should be taken into account when considering the energy contribution of the Coulomb interaction.

The rest of the acceptors together with the hosting semiconductor have been modeled as a free electron bath with
temperature $T$ and chemical potential $\mu_S$. Analogous treatment has been reserved to the tip, to which we have assigned the electrochemical potential
$\mu_0 - eV$ where $e$ is the electron charge taken with sign. Further, we assume that the tunneling between the tip and the foremost acceptors is restricted to the most superficial one ($i = N$). All foremost acceptors are instead coupled to the substrate.
Due to the large distance between the tip and the surface of the semiconductor when compared with the average distance between the acceptors, we have assumed a very asymmetric tunneling coupling of the foremost acceptors to the two ``leads''.

\begin{figure}
\includegraphics{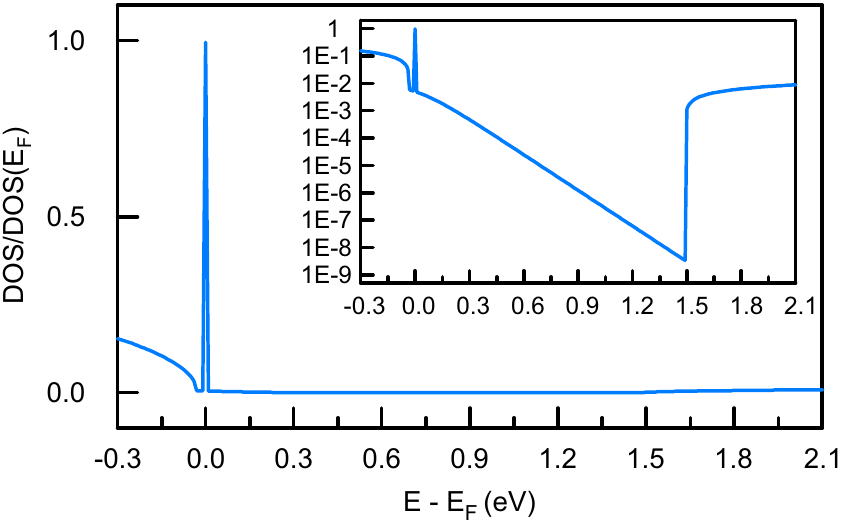}
\caption{Density of states of the substrate used in our model (inset: logarithmic plot).\label{fig5}}
\end{figure}

The dynamics of the system is understood as a sequence of tunneling events from (to) the tip or the substrate which increase (reduce) by one the number of electrons populating the foremost acceptors. The method of choice for the description of this sequential tunneling dynamics is thus the master equation approach. The latter, when applied to second order in the tunneling coupling $H_{\rm tun}$ between the system and the tip (substrate), reduces to the set of rate equations:
\begin{equation}
\label{eq:ME}
\begin{split}
\dot{P}_{ME} &= -\sum_{\chi E'}\large(R^\chi_{ME \to M+1E'}+ R^\chi_{ME \to M-1 E'}\large) P_{ME}\\
&+ \sum_{\chi E'}R^\chi_{M+1 E'\to M E}P_{M+1 E'} + \sum_{\chi E'}R^\chi_{M-1 E'\to M E}P_{M-1 E'}
\end{split}
\end{equation}
where the $P_{ME}$ is the population of the $M$-body eigenstate with energy $E$ of the Hamiltonian $H_{\rm acc }$ given in Eq.~\eqref{eq:Imp_Ham}, and $\chi = S,\,T$ indicates the substrate and the tip, respectively. The many body rates follow from Fermi's golden rule:
\begin{equation}
\label{eq:MB_Rates}
\begin{split}
R^\chi_{ME \to M+1 E'} &= \sum_{\sigma}\sum_{i=1}^{N}\Gamma^\chi_i(E'-E)
|\langle M+1 E'|d^\dagger_{i\sigma}|M E\rangle|^2 f^+(E'-E-\mu_\chi)\\
R^\chi_{ME \to M-1 E'} &= \sum_{\sigma}\sum_{i=1}^{N}\Gamma^{\chi}_i(E-E')
|\langle M-1 E'|d_{i\sigma}|M E\rangle|^2 f^-(E-E'-\mu_\chi)
\end{split}
\end{equation}
where $f^{\pm}(x) = (e^{\pm x/k_{\rm B}T}+1)^{-1}$ are Fermi functions, $\mu_T=\mu_0 - eV$ and $\mu_S = \mu_0$ are the electrochemical potentials of tip and substrate respectively, and $\Gamma^\chi_i$ are the energy dependent single particle tunneling rates defined as:
\begin{equation}
\label{eq:SP_Rates}
\begin{split}
\Gamma^S_i(\Delta E) &= \frac{2\pi}{\hbar}|t_{S}|^2D_S(\Delta E)\\
\Gamma^T_i(\Delta E) &= \frac{2\pi}{\hbar}|t_{T}|^2D_T \delta_{iN}
\end{split}
\end{equation}
where $D_\chi$ is the density of states of the $\chi$ lead which we assume constant for the metallic tip. For what concerns the substrate we have included the energy dependence associated to the impurity band and
the valence and conductance band of the hosting semiconductor.
Fig.~S\ref{fig5} shows a plot of the density of states (DOS) of the substrate, normalized to the DOS at the Fermi energy.
The relative weight of the
three bands are chosen with respect to the density of acceptors and effective masses.
%conduction and valence bands are chosen with respect to the density of acceptors.
We note that we have chosen the DOS within the bulk band-gap to be non-zero ({\it c.~f.} inset of Fig.~S\ref{fig5}),
to %allow for
mimic energy dissipation. Such dissipation can occur via inelastic excitation of vibrons~\cite{Berthe06}.
This DOS drops exponentially with increasing energy, reflecting the expected energy dependence of dissipation efficiency.
We have tested the robustness of our results against different models of the DOS of the substrate.

The substrate tunneling rate is assumed independent of the acceptor index $i$ while the tip tunneling rate is limited to the most superficial acceptor ($i=N$).
%This is justified by the different dimensionality of experiment vs. theory.
Finally the bare tunneling amplitudes $t_{S,T}$ between the leads and the system of foremost acceptors is responsible for the strong asymmetry of the setup yielding $\Gamma^T \ll \Gamma^S$. Since the smallest tunneling rate sets the scale of the current we can estimate $\Gamma^T \approx$~\SI{0.1}{\micro\electronvolt}.% 0.1$~$\mu$ eV.

Notice that in the definition of the substrate many-body tunneling rate (see eq.~\ref{eq:MB_Rates}) we have performed the sum over the acceptor index $i$ only {\it after} taking the square of the many-body matrix element. This %\del{procedure}
approximation destroys any spatial correlation between tunneling events and the associated quantum interference effects.
%\del{We believe this approximation is justified by the many sources of decoherence in the hosting bulk semiconductor.}
In this context, we note that the network of tunneling couplings between the foremost acceptors and the bulk acceptors
is unknown since it depends on the particular acceptor configuration. The approximation applied averages over
all these possible configurations, keeping as the only information an average tunneling rate to the
entire system of foremost impurities.

The $I(V)$ characteristics presented in Fig.~3 of the main text are calculated in the stationary limit. They are obtained by solving first the set of equations (\ref{eq:ME}) in the limit $\dot{P}_{ME} = 0$ and inserting the solution $P^{\rm stat}_{ME}$ in the current formulas:

\begin{equation}
\begin{split}
I_T &= \sum_{MEE'} M \left[-\large(R^T_{ME \to M+1 E'}+ R^T_{ME \to M-1 E'}\large) P_{ME}^{\rm stat}\right.\\
&+ \left. R^T_{M+1 E'\to M E}P_{M+1 E'}^{\rm stat} + R^S_{M-1 E'\to M E}P_{M-1 E'}^{\rm stat})\right]\\
I_S &= \sum_{MEE'} M \left[-\large(R^S_{ME \to M+1 E'}+ R^S_{ME \to M-1 E'}\large) P_{ME}^{\rm stat}\right.\\
&+ \left. R^S_{M+1 E'\to M E}P_{M+1 E'}^{\rm stat} + R^S_{M-1 E'\to M E}P_{M-1 E'}^{\rm stat})\right]\\
\end{split}
\label{eq:current}
\end{equation}
where the (particle) current has been set by convention positive when it increases the number of electrons on the system of foremost acceptors. The stationary limit implies $I_T = -I_S$.

The general expression of the current \eqref{eq:current} is greatly simplified if  $\mu_T \gg E_{N+1} - E_{N}$ for every energy and particle number. This condition is experimentally relevant since it is fulfilled if $V \approx V_{\rm CPD}$. In this high bias limit one readily obtains $R^T_{ME \to M-1E'}\to 0$ and, from \eqref{eq:current}, the current at the tip can be rewritten as:
\begin{equation}
I_T = \sum_{\sigma M E E'}\Gamma_N^T \langle ME| d_{N\sigma} |M+1 E' \rangle \! \langle M+1 E'| d_{N\sigma}^{\dagger} | M E \rangle P^{\rm stat}_{ME}
\end{equation}
Since the states with $M+1$ particle number can be replaced at no prize with a sum over states with a generic particle number $M'$, the current in the high bias limit reads:
\begin{equation}
I_T = \sum_{\sigma M E E'}\Gamma_N^T \langle ME| d_{N\sigma} d_{N\sigma}^{\dagger} | M E \rangle P^{\rm stat}_{ME} = \Gamma_N^T (2 - \langle n_N\rangle)
\end{equation}
where $\langle n_N\rangle$ is the average population of the most superficial acceptor.

In Fig.~3 (main text) we present the simulation of the current as a function of the sample bias for different strength of the interaction parameter $U$. Striking is the appearance, as a function of the interaction strength, of a current plateau around the flat-band condition ($V = V_{\rm CPD}$). This phenomenon is directly linked to the appearance in the system of a new energy scale, i.e. the on-site Coulomb repulsion: in fact the current is essentially pinned to its flat-band value as far as $|{\rm TIBB}(V)| < U/2$.

\begin{figure}[h!]
\includegraphics{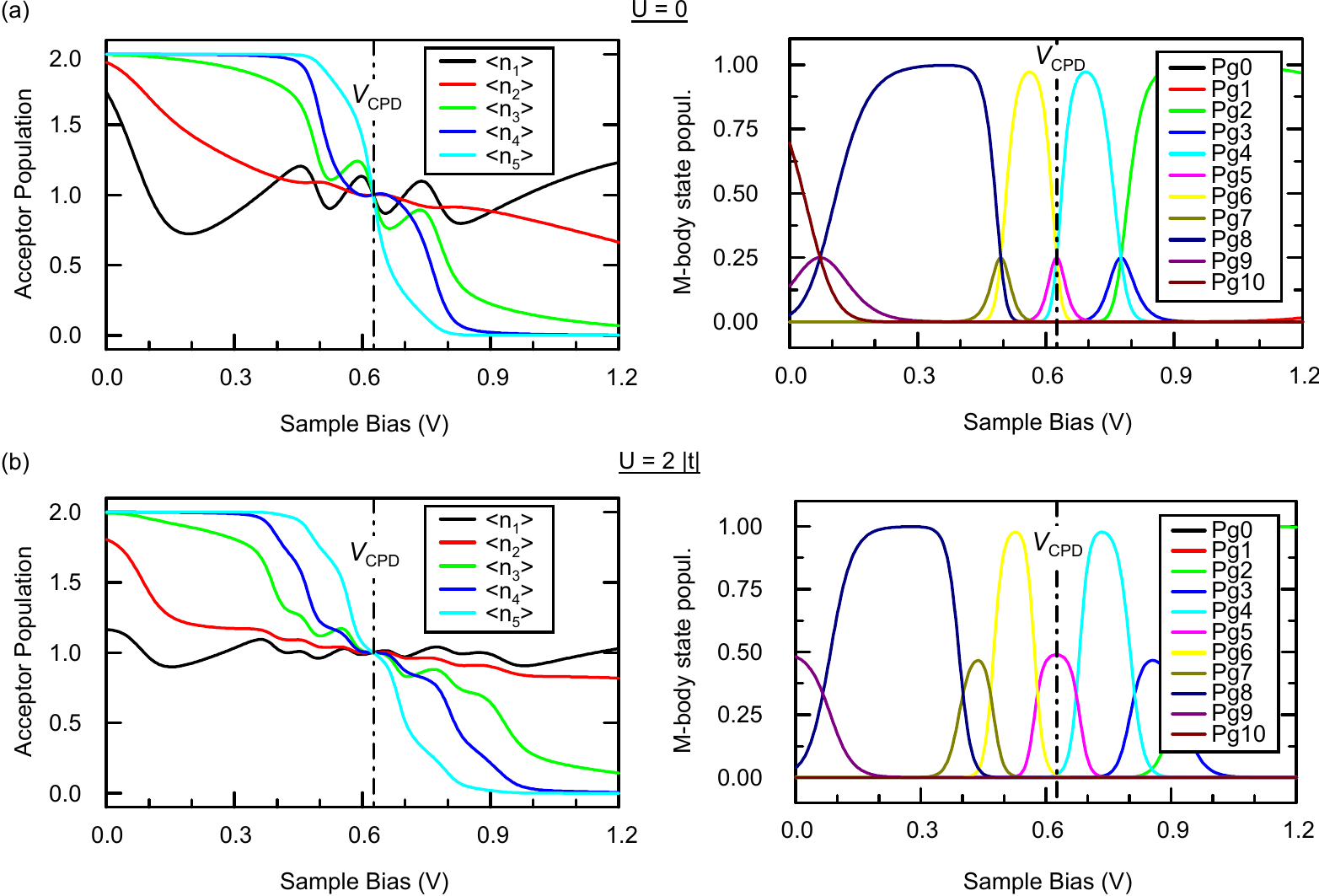}
\caption{(a) Left: Plot of the average occupation $\langle n_i \rangle$ of the 5 foremost acceptors as a function of the sample bias in absence of the on-site Coulomb interaction ($U = 0$). Acceptor 1 is the deepest, acceptor 5 the one closest to the surface. Right: Population of the many-body energy ground states of the 5 foremost acceptors for different particle numbers (Pg), plotted vs. the sample bias. Notice that the ground states with odd particle number are twofold spin degenerate. In each plot, the vertical dashed-dotted line indicates the flat-band voltage.
(b) The same as in (a), but for the case $U = 2|t|$. \label{fig6}}
\end{figure}

A better understanding of this phenomenon can be obtained by comparing the currents plotted in Fig.~3 (main text)
with the acceptors' average occupations and many-body state populations showed in Fig.~S\ref{fig6}.
Both for the vanishing and strong interacting case it is clear that the average occupation of the most superficial acceptor $\langle n_5\rangle$ in the left panels of Fig.~S\ref{fig6} essentially determines the current through the system: the current is blocked whenever the most superficial acceptor is doubly occupied and the more it flows the lower the average acceptor occupation. In fact, if $V > 0$ the particle current flows from the tip to the foremost acceptor to the substrate via tip tunneling events which increase by one the electron number on the most superficial acceptor. These events are Pauli blocked for a doubly occupied acceptor, one transport channel ($|\uparrow\rangle \to |2\rangle$ {\it or} $|\downarrow \rangle \to |2\rangle$) opens if $\langle n_5\rangle = 1$ and 2 channels ($|0\rangle \to |\uparrow\rangle$ {\it and} $|0\rangle \to |\downarrow\rangle$) for $\langle n_5 \rangle= 0$. Finally, it should be noticed that no blocking due to energetic considerations can occur under the approximate flat-band condition $V \approx V_{\rm CPD}$. The system of foremost acceptors is in fact half filled; thus, $E_{N+1 g} - E_{Ng} - \mu_0 \approx U/2$ and a single tip tunneling event can deposit an energy $|eV_{\rm CPD}| \gg U/2$ into the system.

\begin{figure}[h!]
\includegraphics{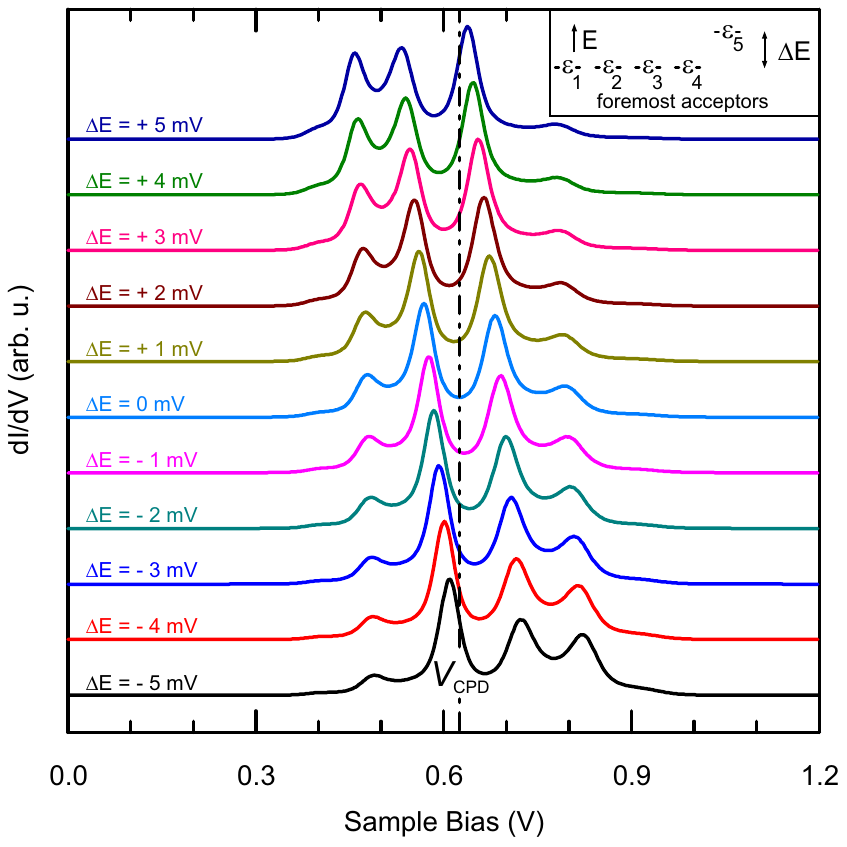}
\caption{\label{fig8} Calculated d$I/$d$V(V)$ spectra, for $U=2|t|$, for eleven different values of the on-site energy of the most superficial acceptor, $\epsilon_5$ (the inset corresponds to TIBB$(V)=0$). The on-site energy is varied in the range $-5$~meV~$\le \epsilon_5 \le +5$~meV. The spectra are vertically offset for clarity. Slight variations of the on-site energy result in considerable shifts of the peak positions with respect to the flat-band voltage $V_{\textrm{CPD}}$ and in changes of the relative peak heights.}
\end{figure}

The average occupation of the last of the foremost acceptors is strongly influenced by the Coulomb repulsion parameter $U$. This influence is best understood by considering the right panels of Fig.~S\ref{fig6}. Firstly one notices that ground state populations dominates the picture even at extremely large biases ($V \approx V_{\rm CPD}$). This is due to the combination of low temperature and strong rate imbalance between the tip and the substrate rates. Essentially we can argue that the foremost acceptors are always {\it in equilibrium} with the substrate. In absence of on-site interaction it does not cost any energy, in the flat-band condition, to add or remove one electron from the half-filling configuration. This is clearly visible in the right panel of Fig.~S\ref{fig6}(a) which shows an equal population of 0.25 for the 4-, 5- (2x spin degenerate) and 6-particle ground states \cite{even-odd}. Still in the non interacting case within TIBB$(V)$ comparable with the impurity bandwidth the 4- (6-) particle ground state reaches full occupation. Most of this change in the particle occupation is concentrating on the last acceptor and correspondingly, on the same energy scale for TIBB$(V)$, the current saturates (vanishes).
\newpage
The interaction lifts this degeneracy and the average population of the half-filled ground state develops a plateau of width $U$ on the scale of TIBB$(V)$, thus approximately $15 |U/e|$ on the bias scale due to level arm scaling. This plateau is readily understood by noticing that the energy needed to add or remove a particle at half filling is exactly $\Delta E = U/2$.

%\del{Finally, we note that we have tested the results of our theoretical model against other values of the number of foremost %acceptors $N$ ( $N=3$, $N=4$), of the on-site Coulomb energy $U$ ($U=0.1|t|$, $U=1|t|$, $U=4|t|$), and the on-site energy $\epsilon_i$~\cite{Wijnheijmer09, Teichmann11} achieving qualitatively similar results.}

Finally, we have tested the results of our theoretical model against variations of different parameters such as: the number of foremost acceptors $N$, the temperature $T$, the on-site energies $\epsilon_i$~\cite{Wijnheijmer09, Teichmann11} and the inter-acceptor tunnelling parameter $t$. The results are reported in Figs.~S\ref{fig8} to S\ref{fig10}.

In Fig.~S\ref{fig8} we show the differential conductance as a function of the bias voltage. The different curves refer to different values of the on-site energy $\epsilon_5$ and are shifted for clarity. While the qualitative picture does not change, even slight variations of the on-site energy (on the order of a few meV) result in considerable shifts of the peak position and relative heights in the differential conductance. This is in accordance with experimental results.

\begin{figure}[h!]
\includegraphics{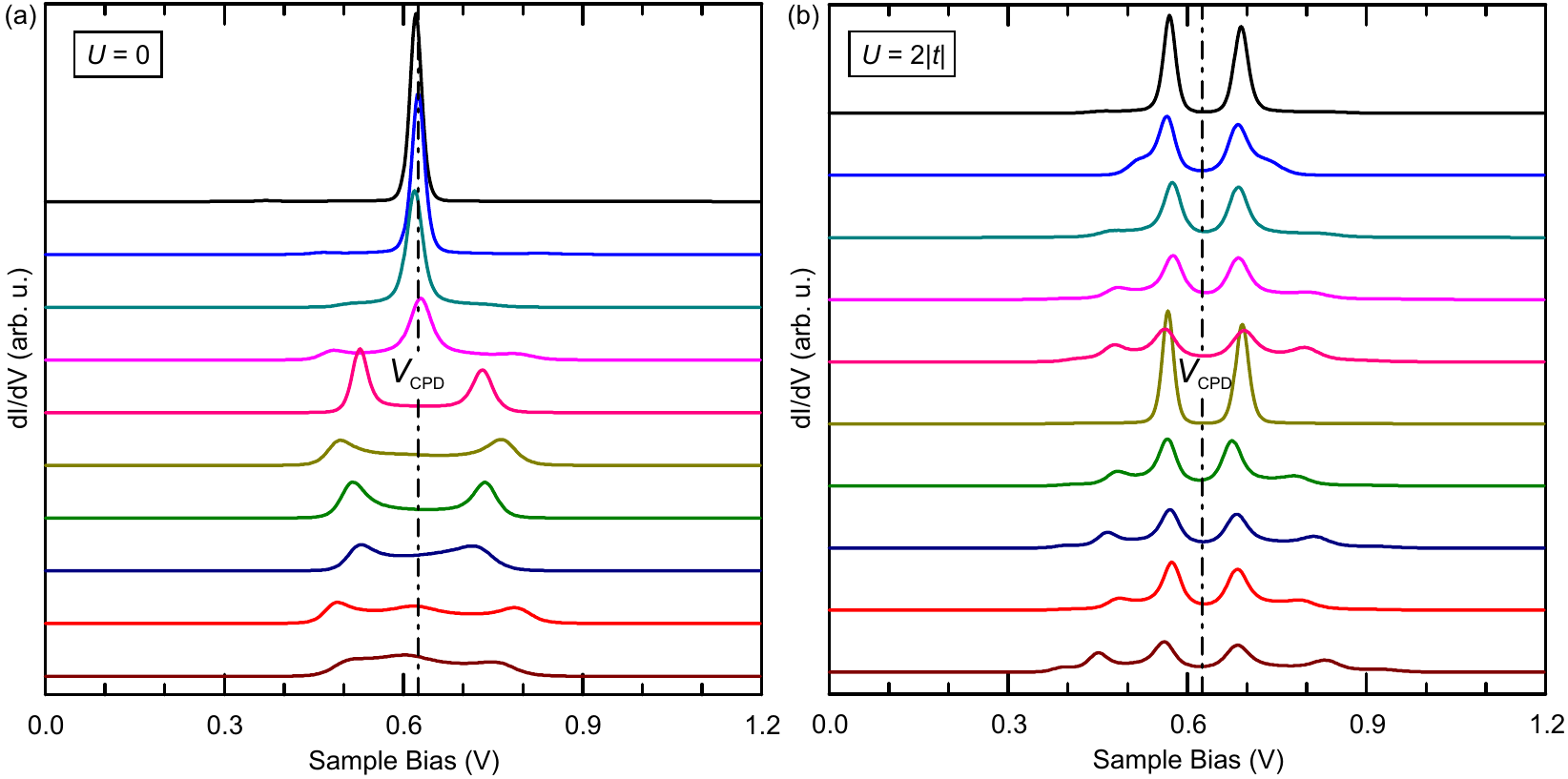}
\caption{\label{fig9} Differential conductance d$I/$d$V(V)$ as a function of sample bias $V$ corresponding to different random realizations of the foremost acceptors Hamiltonian. The different traces are shifted for clarity. Both the on-site energies $\epsilon_i$ and the inter-acceptor tunnelling parameter $t$ are randomized. The  randomization is moderate ( $\epsilon_i \pm 1$~meV) for the on-site energy and exponential (in the range $2$~meV$ - 15$~meV) for the hopping parameter due to the exponential dependence of the tunnelling on the random inter-acceptor distance. In (a) the interaction is switched off while $U = 2|t|$ in (b).}
\end{figure}

In Fig.~S\ref{fig9} we consider instead the effect of the randomization of the on-site energy parameters $\epsilon_i$ and tunneling amplitude $t$ on the transport characteristics of the system. The simulation has been performed both without [Fig.~S9(a)] and with Coulomb interaction [Fig.~S9(b)]. In both cases the randomization produces a large variety of differential conductance traces. In the $U = 0$ case one can distinguish 3 possible scenarios: i) a dominant peak at $V = V_{\rm CPD}$ associated to a weakly coupled last impurity; ii) two well separated peaks, corresponding to a bonding and anti-bonding of the two more strongly hybridized last impurities ; iii) a combination of the two previous pictures. In the finite $U$ case the differential conductance reveals almost always 4 peaks but with strongly varying height and position.

\begin{figure}[h!]
\includegraphics{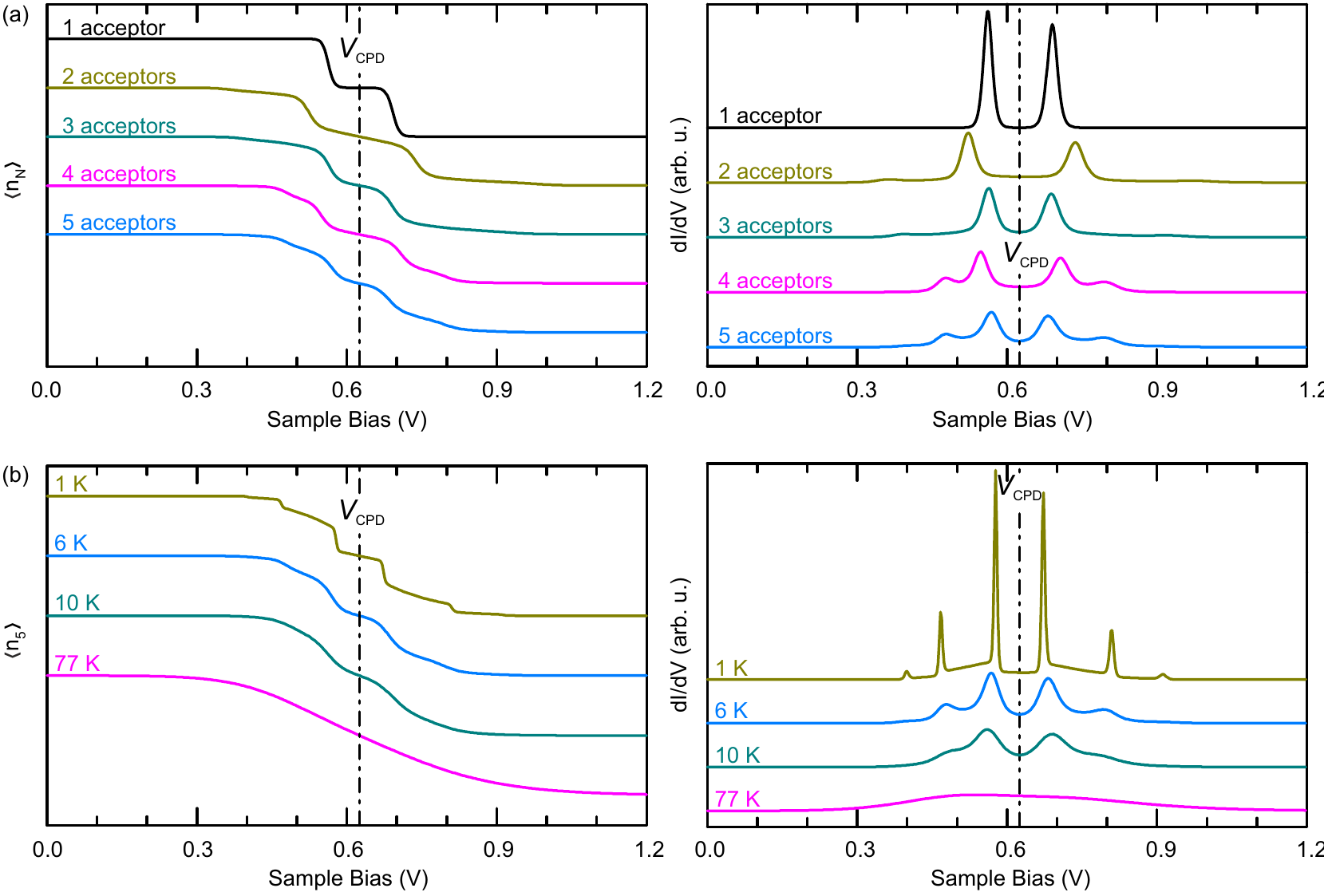}
\caption{\label{fig10} (a) Last impurity population $\langle n_{\textrm N} \rangle$ and differential conductance d$I/$d$V(V)$
as a function of sample bias $V$ for models with a different number of foremost acceptors $N$. Traces are vertically offset for clarity.
(b) shows, for $N=5$, $\langle n_5 \rangle$ and d$I/$d$V(V)$ for different temperatures $T$. Traces are vertically offset for clarity.
}
\end{figure}

In Fig.~S\ref{fig10} we show last impurity population and differential conductance as a function of bias for models with
(a) different number of foremost acceptors, (b) different temperatures.
The variation in the number of foremost impurities produces several effects. Firstly, when passing from the one to the two acceptor model the differential conductance peaks pass from two to four. Their number remains unchanged with increasing impurities ($N = 3, 4, 5$) at least in absence of randomization. Secondly one can appreciate an even-odd effect in the distance between the two peaks closer to the flat band condition: larger distance for even acceptor numbers, smaller for odd ones. This is rationalized by observing that the single particle spectrum for an open chain of $N$ acceptors is given by:

\begin{equation}
E_n = \epsilon + 2 t \cos \left( \frac{\pi n}{N+1}\right)
\end{equation}

\noindent with $n = 1,\,2,\ldots,\, N$. Such a spectrum has a state at the (Fermi) energy $\epsilon$ only for odd number of impurities. In this cases the peak distance reduces to the charging energy (enlarged by the level arm factor). For even $N$'s one should add to the charging energy the distance of the single particle energies from the Fermi level. Thirdly, one notices a progressive closing of the gap around $V = V_{\rm CPD}$ while increasing the number of acceptors. This is the consequence of two facts: firstly a larger number of acceptors decreases the single particle mean level spacing; secondly, the larger number of levels $\epsilon_i$ within the band bending window TIBB$(V)$ increases the delocalization of the foremost impurity states even under strong band bending conditions. This fact, on the other hand, reduces the effective charging energy to pay for the changing in the population of the last impurity associated to the differential conductance peaks. The complete vanishing of this gap would only be possible with a complete delocalization of the impurity states, an absurd condition in presence of strong bend bending. This final analysis reveals the strength of the method which is intrinsically capable to capture the crossover between the delocalized flat band condition and the localized split-off configuration at no price of introducing bias dependent effective parameters.

Finally, the differential conductance illustrated in Fig.~S\ref{fig10}(b) shows the progressive smearing of the spectra with increasing temperatures. The level arm also magnify the temperature broadening and already at $77$ K the conductance peaks are expected to be completely washed out.

%-------------------------------REFERENCES----------------------------------%
\clearpage
\bibliographystyle{apsrev4-1}

\begin{references}
\bibitem{errorKPFM} The standard deviations
%of the parabolic fits
are below 7~mV.
\bibitem{tip_prep}
We note that a non-constant tip density of states may influence experimental $I(V)$ spectra.
To avoid such influence, we prepared metal-terminated apices on Cu(111) before taking spectra on GaAs.
\end{references}
\begin{references}
\bibitem{parameters} The parameters at which KPFS data were acquired on the Cu(111) surface and on the GaAs(110) surface are as follows:
The tip was retracted by
$\Delta z=5$~\AA\, after the feedback-loop of the STM was interrupted at a setpoint of $V=50$~mV, $I=2.5$~pA [on Cu(111)],
or at sample bias ranging from $1.55$~V to $1.8$~V at a current of $I=20$~pA [on GaAs(110)].
\bibitem{error} The standard deviations of the parabolic fits to the Kelvin parabolas are below 4~meV.
\bibitem{paramtip4} For apex~\#4, the tip-sample distance was decreased by 0.25~\AA\,
after the feedback-loop of the STM was interrupted at $V=1.8$~V, $I=20$~pA.
\bibitem{even-odd} For an even number of foremost acceptors the situation would be different. In that case it costs an energy to leave the half filling even in absence of on-site Coulomb repulsion, due to the particular form of the single particle spectrum of an open linear chain. This even/odd effect is though quite irrelevant since the definition of foremost acceptors is somehow arbitrary in the flat-band condition. We choose the odd case which, due to the form of the spectrum (half-filled level at the Fermi energy) better mimics the properties of the infinite ensemble of acceptors. See also Fig.~S10(a) right panel.
\end{references}

\end{document}